\documentclass[twocolumn,aps,prl,amsmath,showpacs]{revtex4}
\usepackage{graphicx}

\begin{document}

\title{Mode-coupling theory for structural and conformational dynamics
of polymer melts}
\author{S.-H. Chong$^{1}$ and M. Fuchs$^{2}$}
\affiliation{$^{1}$Physik-Department, Technische Universit{\"a}t M{\"u}nchen,
85747 Garching, Germany\\
$^{2}$Institut Charles Sadron, 6 rue Boussingault,
67083 Strasbourg, France}
\date{\today}

\begin{abstract}
A mode-coupling theory for dense
polymeric systems is developed which unifyingly incorporates the segmental
cage effect relevant for structural slowing down and polymer chain
conformational degrees of freedom. An ideal glass transition of
polymer melts is predicted which becomes molecular-weight independent
for large molecules. 
The theory provides a microscopic justification for the
use of the Rouse theory in polymer melts, and the results for
Rouse-mode correlators and mean-squared displacements are
in good agreement with computer simulation results.

\end{abstract}


\maketitle

Polymeric materials often find applications as amorphous solids or glasses.
During their processing, the structural relaxation in polymeric
melts plays an important role because it affects transport and viscoelastic
properties. 
An understanding of conformational dynamics is also required for 
single-chain diffusional processes. 
Whereas a number of phenomenological concepts have been invented, 
no microscopic theory exists which successfully explains the 
slow structural and conformational relaxations of polymeric melts
close to the glass transition, as for example studied in detailed
computer simulations of model systems~\cite{Binder01}. 

We present a first-principles theory for structural and conformational
dynamics of unentangled polymer chains using the site formalism which
has already provided a good description of their static 
packing~\cite{Schweizer97}. 
Building upon the mode-coupling 
theory (MCT) for simple and low-molecular 
systems~\cite{Goetze91b,Goetze92,Chong-MCT-dumbbell-1},
our macromolecular extension captures the segmental ``cage effect''
which causes a (asymptotically) molecular-weight-independent ideal
glass transition driven by steric hindrance on microscopic length
scales. 
A decoupling of local collective structural relaxation
from single-chain conformational fluctuations is obtained,
leading to a first-principles derivation of the Rouse model~\cite{Doi86}.
This widely used model assumes 
a chain to be in a Markovian heat bath, and
therefore its applicability
requires a time-scale decoupling as our theory achieves.
Entanglement effects~\cite{Doi86} are neglected in our work because we do not
ensure uncrossability of chains and
consider only local isotropic forces~\cite{Schweizer89}.

A system of $n$ chains, consisting of $N$ identical monomers
or segments, distributed with density $\rho$
is considered.
In the site formalism~\cite{Chandler72}, 
the structural variables are the monomer-density fluctuations
for wave vector ${\vec q}$, 
$\rho_{\vec q}^{a} = \sum_{i=1}^{n} 
\exp(i {\vec q} \cdot {\vec r}_{i}^{\, a})$,
where ${\vec r}_{i}^{\, a}$ denotes the position 
of $a$th monomer in $i$th chain.
Structural dynamics shall be described by 
coherent density correlators, 
$F_{q}^{ab}(t) = 
\langle \rho_{\vec q}^{a}(t)^{*} \rho_{\vec q}^{b} \rangle / n$. 
Here $\langle \cdots \rangle$ denotes canonical averaging for 
temperature $T$. 
A single (or tagged) polymer (labeled $s$) exhibits 
density fluctuations
$\rho_{{\vec q},s}^{a} = \exp(i {\vec q} \cdot {\vec r}_{s}^{\, a})$, and
correlators, 
$F_{q,s}^{ab}(t) = 
\langle \rho_{{\vec q},s}^{a}(t)^{*} \rho_{{\vec q},s}^{b} \rangle$,
characterize the single-chain dynamics. 

We apply MCT equations in the site representation~\cite{Chong-MCT-dumbbell-1}
to flexible macromolecules expecting them to capture
intermolecular caging in polymeric melts. 
The required inputs are the 
static structure factors $S_{q}^{ab} = F_{q}^{ab}(0)$, 
$w_{q}^{ab} = F_{q,s}^{ab}(0)$,
and the direct correlation functions $c_{q}^{ab}$~\cite{Chandler72}. 
There are severe difficulties to solve the $N \times N$-matrix 
MCT equations for polymers because the degrees of polymerization 
$N$ of interest are large. 
The simplifications adopted here are to neglect chain-end effects for
$c_{q}^{ab}$~\cite{Schweizer97}, $c_{q}^{ab} = c_{q}$, and to 
consider the site-averaged correlator,
$F_{q}(t) = (1/N) \sum_{a,b=1}^{N} F_{q}^{ab}(t)$, which
deals with the total monomer-density fluctuations.
This mean-field like approximation replaces the site-specific surroundings of a
segment by an averaged one. 
It is supported by the
observation that $S_{q} = F_{q}(0)$ captures the static
correlations on the segmental length scale~\cite{Schweizer97}. 

We find a set of scalar equations for the normalized coherent correlator 
$\phi_{q}(t) = F_{q}(t) / S_{q}$: 
\begin{eqnarray}
& &
\partial_{t}^{2} \phi_{q}(t) + \Omega_{q}^{2} \phi_{q}(t) 
\nonumber \\
& &
\qquad \qquad + \,
\Omega_{q}^{2} \int_{0}^{t} dt' \,
m_{q}(t-t') \partial_{t'} \phi_{q}(t') = 0,
\label{eq:GLE-phi}
\\
& &
m_{q}(t) = \frac{1}{2} \int d{\vec k} \, V({\vec q}; {\vec k}, {\vec p} \,) \,
\phi_{k}(t) \, \phi_{p}(t).
\label{eq:MCT-phi}
\end{eqnarray}
Here $\Omega_{q}^{2} = q^{2} v^{2} / S_{q}$ and
$V = \rho_{m} S_{q} S_{k} S_{p}
\{ {\vec q} \cdot [{\vec k} c_{k} + {\vec p} c_{p}] \}^{2} / (2
\pi)^{3} q^{4}$
with $v$ denoting the monomer thermal velocity,
$\rho_{m} = N \rho$, and ${\vec p} = {\vec q} - {\vec k}$.
These equations are formally identical to 
the ones for simple systems~\cite{Goetze91b}.
On the other hand, one finds for the single-chain dynamics:
\begin{eqnarray}
& &
\partial_{t}^{2} {\bf F}_{q,s}(t) +
{\bf \Omega}_{q,s}^{2} {\bf F}_{q,s}(t) 
\nonumber \\
& &
\qquad \quad + \,
{\bf \Omega}_{q,s}^{2}
\int_{0}^{t} dt' \,
{\bf m}_{q,s}(t-t') \partial_{t'} {\bf F}_{q,s}(t') = {\bf 0},
\label{eq:GLE-Fs}
\\
& &
m_{q,s}^{ab}(t) =
\sum_{c} \frac{w_{q}^{ac}}{q^{2}}
\int d{\vec k} V_{s}({\vec q}; {\vec k}, {\vec p} \,) \,
F_{k,s}^{cb}(t) \, \phi_{p}(t).
\label{eq:MCT-Fs}
\end{eqnarray}
Here 
${\bf \Omega}_{q,s}^{2} = q^{2} v^{2} {\bf w}_{q}^{-1}$ and
$V_{s} = 
\rho_{m}
({\vec q} \cdot {\vec p} / q)^{2} \, S_{p} c_{p}^{2} / (2\pi)^{3}$.

A traditional description of the single-chain dynamics 
is in terms of Rouse modes~\cite{Doi86}. 
For discrete chains, Rouse-mode correlators are defined by
$C_{pp'}(t) = \langle {\vec X}_{p}(t) \cdot {\vec X}_{p'} \rangle / 3N$
with
${\vec X}_{p} = \sqrt{2/N} 
\sum_{a=1}^{N} {\vec r}_{s}^{\, a}
\cos[(a-1/2)p \pi/N]$. 
Since $\rho_{{\vec q},s}^{a} \approx 
1 + i {\vec q} \cdot {\vec r}_{s}^{\, a}$ for small ${\vec q}$, 
$C_{pp'}(t)$ can be expressed
as a linear combination of $F_{q,s}^{ab}(t)$ for $q \to 0$. 
Here, it is neither assumed that $C_{pp'}(t)$
are diagonal, nor that the decay is exponential. 
Therefore, our microscopic theory can test 
the validity of the Rouse theory. 

The polymers considered here shall be modeled as follows. 
First, chains are assumed to be Gaussian, for which 
$w_{q}^{ac} = \exp[ -q^{2} | a - c |\, b^{2}/6]$
with the statistical segment length $b$. 
Second, each monomer is modeled as a hard sphere of diameter $d$
and mass $m$,
and we set $b = d$.
All equilibrium properties are then specified by the
packing fraction
$\varphi = \pi \rho_{m} d^{3} / 6$ and the degree of polymerization
$N$. 
Third, $S_{q}$ and $c_{q}$ are evaluated from the 
polymer RISM theory~\cite{Schweizer97}.
Let us add that the value of $\varphi$ can become larger than 1 for very 
dense systems since corrections due to unphysical intrapolymer monomer 
overlap~\cite{Schweizer88} are not taken into account in the present work.
From here on, 
the units will be chosen so that $d=v=m=1$. 

\begin{figure}
\includegraphics[bb=70 505 344 724,totalheight=5.5cm,keepaspectratio]{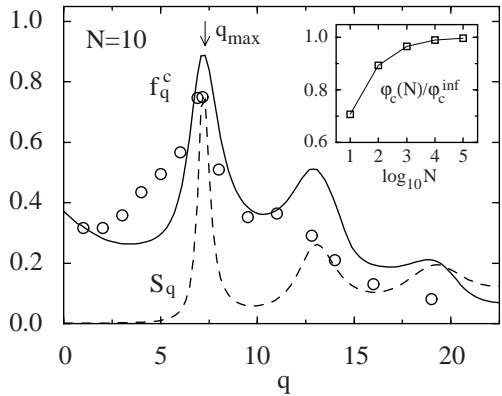}
\caption{
The solid line exhibits
the glass-form factors $f_{q}^{c}$ 
at $\varphi = \varphi_{c}$
for $N=10$ along with the simulation
result (circles) for a similar system
taken from Ref.~\onlinecite{Aichele01}.
The dashed line denotes $S_{q}$ multiplied by 0.15. 
The inset shows $\varphi_{c}(N)$ rescaled by
$\varphi_{c}^{\rm inf} = \varphi_{c}(N \to \infty) \approx 1.309$ 
as a function of $\log_{10} N$.}
\label{fig:fqc-Sq-phic}
\end{figure}

The derived MCT equations (\ref{eq:GLE-phi}) and (\ref{eq:MCT-phi}) 
exhibit an ideal liquid-glass transition upon
increasing $\varphi$ at a critical value $\varphi_{c}$
as discussed for simple systems~\cite{Goetze92,Franosch97}. 
The solid line in Fig.~\ref{fig:fqc-Sq-phic} exhibits
the resulting glass-form factors 
$f_{q}^{c} = \phi_{q}(t \to \infty)$ 
at $\varphi_{c}$ for $N=10$, and the
dashed line $S_{q}$ multiplied by 0.15.
The $f_{q}^{c}$ also measure the amplitude
of the slow structural relaxation in the liquid state. 
Inspection of Eq.~(\ref{eq:MCT-phi}) shows that
the liquid-glass transition is driven mainly by the
changes of $S_{q}$ for $q_{\rm max} \approx 7$ where $f^c_q$ is maximal, 
i.e., 
by local fluctuations connected to the average
nearest-neighbor-monomer distance (the cage effect). 
With varying $q$, $f_{q}^{c}$ oscillates in phase with 
$S_{q}$ as found in simple systems~\cite{Goetze92,Franosch97}. 
The circles denote the result of molecular-dynamics simulation
performed for a similar 
system~\cite{Bennemann98,Bennemann99,Bennemann99b,Aichele01}.
The agreement is semiquantitative especially for $q \approx q_{\rm max}$,
which is the relevant $q$ range for the ideal glass transition
\cite{disagreementnote}. 
The inset demonstrates
that $\varphi_{c}(N)$ becomes independent of $N$ for large $N$.
This is because the glassy arrest is driven by the local cage
effect, and the global chain size plays a gradually
smaller role for larger $N$~\cite{McKenna89}. 

\begin{figure}
\includegraphics[bb=70 446 396 724,totalheight=6cm,keepaspectratio]{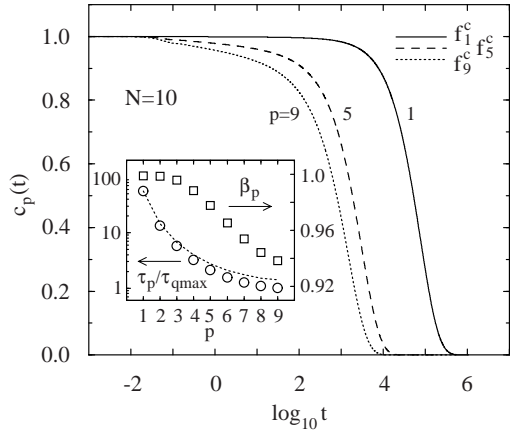}
\caption{
Rouse-mode correlators
$c_{p}(t)$ of $p=1$ (solid), $p=5$ (dashed), and
$p=9$ (dotted lines) for $N = 10$
and $(\varphi - \varphi_{c})/\varphi_{c} = - 10^{-2}$. 
The horizontal lines mark the plateaus $f_{p}^{c}$. 
The inset shows the ratio $\tau_{p}/\tau_{\rm qmax}$ of 
the $\alpha$-relaxation times (circles) and the
stretching exponent $\beta_{p}$ (squares)
as a function of $p$.
The dotted line denotes a fitting based on the formula
${\tau}_{p} = \zeta / 12 \sin^{2}(p \pi / 2N)$.}
\label{fig:Rouse-correlators}
\end{figure}

For packing fractions $\varphi$ close to but below $\varphi_{c}$, 
correlator $\phi_{A}(t)$ of any variable $A$ 
coupling to density fluctuations exhibits slow structural relaxation and
decays in two steps: the decay towards the plateau
$f_{A}^{c}$, followed by the decay to zero 
(so-called $\alpha$-decay)~\cite{Goetze91b,Goetze92}.
Detailed analysis of $\phi_{q}(t)$ and $F_{q,s}^{ab}(t)$ 
will be presented in later publications. 
Here, only the Rouse-mode correlators $C_{pp'}(t)$ shall be studied.
Typically, their off-diagonal elements are found to be much smaller 
(at most only a few \%) compared to the diagonal ones at all times. 
Therefore, the normalized diagonal elements,
$c_{p}(t) = C_{pp}(t)/C_{pp}(0)$, shall be considered. 
Figure~\ref{fig:Rouse-correlators} exhibits representative
results for $N = 10$ 
for a reduced packing fraction
$(\varphi - \varphi_{c})/\varphi_{c} = - 10^{-2}$. 

The curves shown in Fig.~\ref{fig:Rouse-correlators}
do not clearly exhibit the two-step-relaxation scenario.
This is because the plateaus of $c_{p}(t)$
are so large, $f_{p}^{c} > 0.9$, 
that only less than 10\% of the decay are left for the
relaxation towards the plateau.
Thus, most of the relaxation of $c_{p}(t)$ occurs
in the $\alpha$ regime.
Let us characterize the $\alpha$-time scale 
by $c_{p}(\tau_{p}) = f_{p}^{c}/20$.
The corresponding $\alpha$-time scale $\tau_{\rm qmax}$ shall 
be introduced for the coherent correlator $\phi_{q}(t)$
for $q = q_{\rm max}$, characterizing 
the local dynamics of the surrounding medium. 
The ratio $\tau_{p} / \tau_{\rm qmax}$ is shown in the inset
as a function of $p$. 
For small $p$, the scales $\tau_{p}$
are much larger than the local scale $\tau_{\rm qmax}$. 

The Rouse theory assumes that
all dynamical correlations in the surroundings
are much faster than the single-chain dynamics~\cite{Doi86}.
Since a polymer is surrounded 
by identical polymers, 
the assumption of the time-scale separation cannot be justified
{\it a priori}. 
Our microscopic theory verifies this central assumption. 
The Rouse theory predicts, within our units,
$\tau_{p} = \zeta / 12 \sin^{2}(p \pi / 2N)$
where $\zeta$ denotes the monomer friction.
This formula, using $\zeta$ as a fit parameter,
is shown as the dotted line, and 
we found $\zeta / \tau_{\rm qmax} = 16.5$. 

The shape of $c_{p}(t)$ in the $\alpha$ regime
is often characterized by the stretching exponent $\beta_{p}$ of the
Kohlrausch-law fit: 
$c_{p}(t) \propto \exp[ - (t/\tau_{p}^{\prime})^{\beta_{p}} ]$.
We found $\beta_{p} > 0.9$ for all $p$ 
as shown in the inset. 
For small $p$, $\beta_{p}$ is close to 1 due to the large
separation of the scales $\tau_{p}$ and $\tau_{\rm qmax}$
as discussed above.
$\beta_{p}$ decreases as $p$ increases
since for large $p$ the scales $\tau_{p}$ become
comparable to $\tau_{\rm qmax}$.
However, $\beta_{p}$ remains close to 1
even for large $p$.
This is because their plateaus $f_{p}^{c}$ are high
as explained in Ref.~\onlinecite{Goetze00c}.
Thus, all our $c_{p}(t)$ 
exhibit nearly Debye-relaxation as assumed by the 
Rouse theory.
Let us note that the found features for $c_{p}(t)$ 
hold also for $N=100$, and are in agreement with simulation
results~\cite{Bennemann99}. 

\begin{figure}
\includegraphics[bb=70 176 396 724,totalheight=12cm,keepaspectratio]{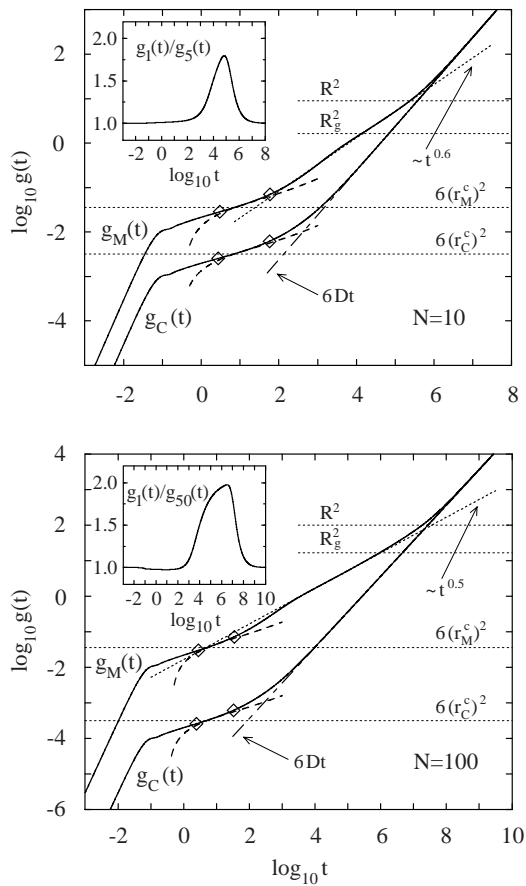}
\caption{
Double logarithmic presentation of the MSDs
$g_{M}(t)$ and $g_{C}(t)$ (solid lines) 
for $N=10$ (upper) and 100 (lower panel) 
for $(\varphi - \varphi_{c})/\varphi_{c} = - 10^{-2}$. 
The dashed lines denote the asymptotic formula
given in Eq.~(\protect\ref{eq:MSD-first}).
The open diamonds mark the points where the dashed lines differ from 
the solid ones by 10\%.
The dashed-dotted lines exhibit the diffusion law, $6Dt$.
The dotted lines show the power laws, $\sim t^{x}$, with 
$x$ specified in each panel.
The horizontal dotted lines from bottom to top successively mark the 
plateau $6 (r_{C}^{c})^{2}$ for $g_{C}(t)$, 
the plateau $6 (r_{M}^{c})^{2}$ for $g_{M}(t)$, 
the radius of gyration $R_{g}^{2}$, and the end-to-end distance $R^{2}$.
The inset exhibits the ratio $g_{1}(t)/g_{N/2}(t)$.}
\label{fig:MSD}
\end{figure}

Of interest are the monomer mean-squared
displacement (MSD) 
$g_{a}(t) = \langle [ {\vec r}_{s}^{\, a}(t) - {\vec r}_{s}^{\,
a}(0)]^{2} \rangle$ and the center-of-mass MSD, $g_{C}(t)$.
Let us also introduce the monomer-averaged MSD,
$g_{M}(t) = (1/N) \sum_{a=1}^{N} g_{a}(t)$.
Figure~\ref{fig:MSD} exhibits $g_{M}(t)$ and $g_{C}(t)$
for $N=10$ and 100
for $(\varphi - \varphi_{c})/\varphi_{c} = - 10^{-2}$. 
For short times, the MSD exhibits ballistic motion:
$g_{X}(t) \propto t^{2}$ ($X=M$ or $C$). 
As the time increases, the MSD begins to be suppressed due to the
cage effect, and there appears the so-called $\beta$ regime
where $g_{X}(t)$ 
is close to the square of the critical localization
length, $g_{X}(t) \approx 6 (r_{X}^{c})^{2}$. 
For this regime, there holds~\cite{Goetze91b,Goetze92,Fuchs98}
\begin{equation}
g_{X}(t) = 6 (r_{X}^{c})^{2} - 6 h_{X} G(t), \quad
X = M, C,
\label{eq:MSD-first}
\end{equation}
where $G(t)$ denotes the $\beta$ correlator
and $h_{X}$ the critical amplitude. 
Thus, $g_{M}(t)$ and $g_{C}(t)$ cross
their plateaus $6 (r_{X}^{c})^{2}$ at the same time.
The value $r_{M}^{c} \approx 0.077$,
quantifying the monomer localization, 
for both $N=10$ and 100
is consistent with Lindemann's melting criterion.
$r_{C}^{c}$ is reduced by about $1/\sqrt{N}$
compared to $r_{M}^{c}$, as expected for independent motions
of constituent monomers. 
The asymptotic law (\ref{eq:MSD-first}) for each MSD
is drawn as dashed line, and
its range of validity is indicated by diamonds. 
The trends in $g_{X}(t)$ up to the $\beta$ regime
are qualitatively similar to the ones for a sphere 
in a simple system~\cite{Fuchs98}. 

The increase of $g_{X}(t)$ above the plateau $6 (r_{X}^{c})^{2}$
towards the diffusion asymptote,
$g_{X}(t) \approx 6 Dt$ with the diffusivity $D$, 
is the $\alpha$ process of the MSD. 
In contrast to $g_{C}(t)$, 
$g_{M}(t)$ in this regime 
is significantly affected by  chain connectivity
since the segments participate in the conformational motion and
most of the relaxation of $c_{p}(t)$ occurs here
as explained above.
As a result, 
there appears a subdiffusive ($\sim t^{x}$) regime in $g_{M}(t)$.
For $N=10$, the exponent $x$ is 0.60,
which is close to the value (0.63) found in the mentioned
simulations~\cite{Bennemann99}. 
For $N=100$, we find $x = 0.5$ as predicted by the
Rouse model and by the asymptotic evaluation of our theory, and
the different value for $N=10$ can be attributed to finite $N$ effects. 
Thus, we find a strong polymer-specific effect for the beginning of the
$\alpha$ process, while no such effect
is reflected in $g_{M}(t)$ up to the $\beta$ process.

The insets of Fig.~\ref{fig:MSD} show the ratio of the 
MSD for end and central monomers, $g_{1}(t) / g_{N/2}(t)$.
The ratio starts from 1 in the ballistic regime, 
exhibits a maximum for intermediate times, and tends to 1 
in the diffusion regime.
It is seen that the ratio remains close to 1 also 
within the $\beta$ regime.
This is because the dynamics here does not reflect the 
chain connectivity as explained above.
The Rouse theory predicts the maximum to be 2
within the time regime where the monomer MSD exhibits the 
$t^{1/2}$ law, i.e. in the $\alpha$ regime.
The result for $N=100$ clearly indicates this behavior,
while the maximum for $N=10$ is slightly 
smaller (1.8) due to finite $N$ effects. 
The shape of the ratio for $N=10$
in the $\beta$ and $\alpha$ regimes is 
in semiquantitative agreement with the mentioned
simulation results~\cite{Baschnagel-private}.
Thus, the chain-end effect is properly taken into account in our 
theory even with neglecting that effect for 
$c_{q}^{ab}$. 
This is because the matrix structure of 
Eqs.~(\ref{eq:GLE-Fs}) and (\ref{eq:MCT-Fs})
are preserved for describing the single-polymer dynamics. 

In summary, a first-principles theory for structural slowing down
of dense polymer systems has been presented which also provides a microscopic
derivation of the Rouse model for unentangled chain melts. Chain
connectivity is seen to cause  polymer specific long-time anomalies
of the $\alpha$-process.
 We use the concept of an ideal MCT glass transition, familiar
for simple liquids and colloidal suspensions~\cite{Goetze99}, and
our results agree well with simulation studies of simple (coarse
grained) polymer models. 
Our theory thus provides insights into dynamical aspects typical for
polymer melts
which are also observed in models~\cite{Zon98,Zon99}
with a more realistic local chemistry.

\begin{acknowledgments}
We thank J. Baschnagel, M. Aichele, and W. G\"otze for discussions.
M.F.\ was supported by the Deutsche Forschungsgemeinschaft, grant Fu~309/3. 
\end{acknowledgments}


\begin{thebibliography}{22}
\bibitem[{dis()}]{disagreementnote}
\bibinfo{note}{The disagreement for $q < q_{\rm max}$ may be due to the fact
  that center-of-mass density fluctuations are not our explicit dynamical
  variables~\protect\cite{Aichele01}, while the one for $q > q_{\rm max}$ may
  be due to softer Lennard-Jones potential which is employed in the
  simulation.}
\end{thebibliography}
\end{document}